\pdfoutput=1

\documentclass[11pt]{article}

\usepackage{acl}

\usepackage{times}
\usepackage{latexsym}

\usepackage[T1]{fontenc}

\usepackage[utf8]{inputenc}

\usepackage{microtype}
\usepackage{graphicx}
\usepackage{subfigure}
\usepackage{booktabs}
\usepackage{threeparttable}
\usepackage{xspace}

\AtBeginDocument{%
  \providecommand\BibTeX{{%
    \normalfont B\kern-0.5em{\scshape i\kern-0.25em b}\kern-0.8em\TeX}}}

\usepackage{amsmath,amsfonts,algorithm}
\usepackage[noend]{algpseudocode}

\title{Training a Tokenizer for Free with Private Federated Learning}

\author{%
Eugene Bagdasaryan%
\thanks{~~Work done during the internship at Apple.}
 \\ Cornell Tech \\ \texttt{eugene@cs.cornell.edu} \\\AND
Congzheng Song \and
Rogier van Dalen \and
Matt Seigel \and
\'{A}ine Cahill \\ Apple \\ \texttt{\{csong4,rogier\_vandalen,mseigel,aine\_cahill\}@apple.com}
\\}

\begin{document}
\maketitle

\newcommand{\paragraphbe}[1]{\vspace{0.75ex}\noindent{\bf \em #1}\hspace*{.3em}}
\newcommand{\eb}[1]{{\textcolor{blue}{[EB: #1]}}}
\newcommand{\BOS}{\texttt{BOS}} 
\newcommand{\EOS}{\texttt{EOS}}
\newcommand{\OOV}{\texttt{OOV}\xspace}


\begin{abstract}


Federated learning with differential privacy, i.e.\ private federated
learning (PFL), makes it possible to train models on private data
distributed across users' devices without harming privacy.
PFL is efficient for models, such as neural networks, that
have a fixed number of parameters, and thus a fixed-dimensional gradient
vector.
Such models include neural-net language models, but not tokenizers, the topic of this work.
Training a tokenizer requires frequencies of words from an unlimited vocabulary, and existing methods for finding an unlimited vocabulary need a separate privacy budget.

A workaround is to train the tokenizer on publicly available data.
However, in this paper we first show that a tokenizer trained on mismatched data results in worse model performance compared to a privacy-violating ``oracle''
tokenizer that accesses user data, with perplexity increasing by 20\,\%.
We also show that sub-word tokenizers are better suited to the federated context than word-level ones, since they can encode new words, though with more tokens per word.

Second, we propose a novel method to obtain a tokenizer without using any additional privacy budget.
During private federated learning of the language model, we sample from
the model, train a new tokenizer on the sampled sequences, and update
the model embeddings.
We then continue private federated learning, and obtain performance within 1\,\% of the ``oracle'' tokenizer.
Since this process trains the tokenizer only indirectly on private data, we can use the ``postprocessing guarantee'' of differential privacy and thus use no additional privacy budget.

\end{abstract}


\section{Introduction}

Learning a language model (LM) requires text data that in many
situations is private, resides on people's devices, and should stay
there. In federated learning \citep{fedlearn_1}, a central server learns
a model by receiving statistics, like parameter updates, from many
devices.  Though devices send only statistics and not the raw data,
federated learning by itself can leak information about the data
\citep{shokri2017membership,song2017machine}. Private federated learning
(PFL) \cite{fedlearn_dp, geyer2017differentially} uses differential
privacy \citep{dwork2006calibrating,dwork2014algorithmic} to mitigate
the privacy leaks by limiting the user's impact on the final model.

It is known how to train neural-net language models using PFL
\citep{fedlearn_dp}.  However, an important part of language modeling is
tokenization: turning a text into a sequence of symbols from a fixed-size
symbol set.  To obtain a tokenizer, published research on private
federated learning of language models uses either of two approaches,
neither of which are satisfactory.  One approach is to train the
tokenizer on user data directly.  The commonly-used LEAF dataset
\cite{caldas2018leaf} and works relying on it \cite{li2021ditto,
hu2021private, yu2020salvaging} assume access to the training data to
create the tokenizer.  This is not relevant to real-world use cases and
undermines user privacy.  The other approach is to use public
data to obtain the tokenizer \cite{fedlearn_dp}.  This is sensible from
a privacy perspective, but as we show the resulting distribution
mismatch harms performance, resulting in 10\%-20\% drop compared to
using an ``oracle'' tokenizer trained directly on users' private data. 

\begin{figure}[t]
    \centering
    \includegraphics{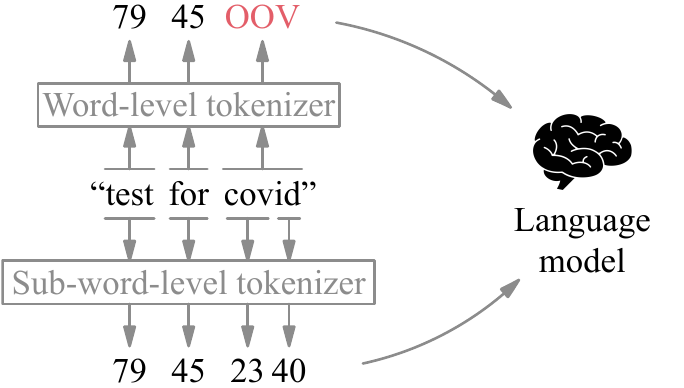}
    \caption{Word-level and sub-word-level tokenization.
        A word-level tokenizer can generate an ``out-of-vocabulary'' (OOV) symbol, which it is hard for a language model to use.
        \label{fig:word_sub-word}}
\end{figure}

There are two common types of tokenization, which are affected by
mismatched distributions in different ways: word and sub-word
tokenization.
Figure \ref{fig:word_sub-word} illustrates these.
A word-level tokenizer produces a symbol for each word, and assigns an out-of-vocabulary
token (OOV) to any unseen word.  Text from mismatched distributions
will generally contain unseen words, which means the correct word cannot
be predicted, and the context becomes less meaningful when predicting
the next word.
Sub-word tokenization, on the other hand, splits some words into multiple
smaller tokens.  This type of tokenization is generally chosen to
minimize the average number of tokens per word on training data.  Current centrally
trained models use sub-word tokenization such as Byte-Pair
Encoding~\cite{sennrich2016neural},
SentencePiece~\cite{kudo2018sentencepiece}, or
WordPieces~\cite{schuster2012japanese}. Nevertheless, mismatched
tokenizations in sub-word methods cause an increase in the number of
tokens per word, and thus decrease the amount of context the model can
use to predict the distribution of the next word.

In this work we present a general framework to approach training
language models in private federated learning by including tokenization
as part of the training pipeline. Our contributions are: (1) we uncover
the performance gaps when the models use the tokenizer obtained from a
different distribution vs the tokenizer obtained from the underlying
distribution. For word-level tokenization we show that a tokenizer
trained on public data reduces the next-word prediction accuracy of
10--20\,\% compared to a tokenizer estimated on user data.  (2) We
demonstrate significant benefits of switching tokenizers from word to
sub-word level, thus eliminating the out-of-vocabulary problem.  (3) We
propose a new method that samples data from an existing model, e.g. from
the prior PFL run, and uses that data to initialize a new tokenizer.
Our approach can update the tokenizer between iterations of
the same PFL run by modifying model embeddings with new tokenizations and
significantly boosting performance.
Crucially, since the language model is trained with differential privacy, the ``postprocessing guarantee'' of differential privacy means that training the tokenizer with our approach does not use any additional privacy budget.

\section{Private federated learning}

Machine-learned models work best if they are trained on the correct distribution of the data, in this paper text data.
In many scenarios text data is private and contained on people's devices, and should stay there.
To train a global model without harming privacy, we use federated learning \citep{fedlearn_1} with differential privacy \cite{dwork2006calibrating,dwork2014algorithmic}.

Federated learning involves devices sending not the data, but statistics, e.g.\ model gradients, computed on that data.  
To train neural networks, the standard algorithm is \emph{federated averaging} \citep{fedlearn_1}.
At each iteration $t$, the server randomly selects a subset of $m$ participants $S_m$ and distributes the current global model $M^t$.
Each participant takes a number of gradient steps to train on their private
data and submits the sum $G_i^t$ of the gradients to the server.
The server takes a step (with step size $\eta$) in the direction of the average gradient to create the new global model:
\begin{equation}
    \label{eq:fed_avg}
    M^{t+1} = M^{t} + \frac{\eta}{m}\sum_{i=1}^m G_i^t
\end{equation}

\subsection{Federated Learning with Differential Privacy}

The global model $M^{t+1}$ might still reveal private
information including user participation in
training \citep{shokri2017membership,song2017machine,melis2018inference}.
To mitigate this threat, we can combine federated learning with
differential privacy (DP)
\citep{dwork2006calibrating,dwork2014algorithmic}, to give \emph{private
federate learning} \citep{fedlearn_dp}.  Differential privacy gives a
strong guarantee: it limits the advantage that a computationally
unconstrained adversary has in inferring whether an individual's data is
contained in the data set that the statistics are computed from.
$(\epsilon, \delta)$-differential privacy parametrizes this advantage by
$\epsilon$ (the maximum privacy loss) and $\delta$ (a slack term).  The
common mechanism to provide differential privacy in a federated learning setting
is the Gaussian mechanism that uses the \emph{moments
accountant} \citep{abadi2016deep}. For each participant, the model parameters are
\emph{clipped} to a norm $S$, i.e., multiplied by $\textnormal{min} (1,
S/{\lVert G^t\rVert_2})$, to bound the sum's sensitivity to any individual's data.
Second, Gaussian noise $\mathcal{N}(0,\sigma^2)$ is added to the final sum.
How much privacy budget is spent depends on the variance $\sigma^2$ relative to the magnitude of individual updates, the total population, the number of contributions in each iteration, and the total number of iterations \citep[for more details, see][]{fedlearn_dp,borja2018subsampling}.

\subsection{Privately finding vocabulary items}


Central differential privacy with the Gaussian mechanism and the moments accountant is efficient in terms of utility vs privacy loss, but it does come with restrictions.
The sum of individual contributions, which the noise is added to, must be of finite and fixed size.
This is not a problem for training neural networks.
However, training a tokenizer requires frequencies for an exponential-size set of sequences, as does training a traditional $N$-gram model.
Differentially private algorithms to compute histograms over sets of elements (e.g.\ words) distributed over devices are
called ``heavy hitters'' algorithms
\citep{bassily2017practical,zhu2020federated,apple2017learning}.
These algorithms require a separate and large privacy budget.
In section~\ref{sec:exps} we will compare with a heavy hitters algorithm.

Another way of finding vocabulary items privately is to train a
neural-net generative model. \Citet{beaufays2019oov} trains a separate,
character-level LSTM model to generate the new words. However, the
proposed method is only shown to work for discover {\OOV}s in a word-level model and
also requires separate training and a privacy budget. 


\section{Tokenization in Language Modeling}
\label{sec:tokenization}

A language model is a model that assigns
probabilities to sequences of tokens.  In this paper, it is always an
autoregressive model with parameters $\theta$: $ P_\theta(s) =
P_\theta(t_2|t_1=\BOS) \cdot P_\theta(t_3|t_1=\BOS, t_2) \cdots
P_\theta(t_n=\EOS | t_1=\BOS, \ldots, t_{n-1}) $, where each term in
this equation is normalized over all possible values of the current
token.  Local normalization is useful when decoding input, like in
speech recognition or a keyboard \cite{hard2018federated}.
For this paper, we assume that a corpus is segmented into sentences.  A
tokenizer $\tau$ then converts each sentence $s$ in the dataset into a
sequence of $n$ tokens $\tau(s) = [\BOS, t_2, .., t_{n-1}, \EOS]$, which is fed into the language model.
There are two types of tokenization, highlighted in Figure \ref{fig:word_sub-word}: word-level and sub-word-level.
Using a sub-word tokenizer will be key to the algorithm this paper proposes.

The next section will discuss the two types of tokenizers and their consequences for out-of-vocabulary tokens and the performance of language models based in them.
Section \ref{sec:compare_tokenizations} will discuss the complex topic of how to compare performance across different tokenizations.

\subsection{Word-level vs sub-word-level tokenization}

The type of tokenization that papers about language models in federated learning commonly use is
word-level tokenization~\cite{fedlearn_1}. For a vocabulary of size $N$
the tokenizer assigns a unique token for top-$N$ most popular words in
the dataset while other words receive an out-of-vocabulary token {\OOV}, as highlighted in Figure \ref{fig:word_sub-word}.
Some papers \citep[e.g.][]{fedlearn_dp} build the tokenizer from a
publicly available dataset, others including the LEAF benchmark
\cite{caldas2018leaf} build the tokenizer from users' training data.
OOV tokens in the word history make it harder for a language model to predict the next word.

The other type of tokenization is sub-word tokenization, for which there are two popular schemes: byte-pair
encoding (BPE) \cite{sennrich2016neural}
and WordPieces \citep{schuster2012japanese}.  We focus on BPE which
unlike WordPieces guarantees the absence of OOVs as there exists a token for every byte.
However, the number of tokens required to encode each word can change significantly depending on the dataset that the tokenizer was trained on.
As highlighted in Figure \ref{fig:word_sub-word}, a tokenizer trained on data from before the COVID-19 pandemic would generate multiple tokens for the word ``covid''.

Generating longer token sequences makes it harder for the language model to keep track of the context, degrading its performance.
Even LSTMs and transformers, which in theory can use arbitrarily long history,
have imperfect memory.

\subsection{Evaluating language models across tokenizations}
\label{sec:compare_tokenizations}

Comparing language models across tokenizations is a complex problem. For
example, when comparing word-level language models using perplexity,
often OOVs are ignored which gives an edge to the language model with
more OOVs, which is the opposite of what is desired.  The following
sections detail the problems when comparing sub-word language models.

\subsubsection{Comparing word-level with sub-word}
Since a word-level language model has a closed vocabulary, it outputs
probabilities only on in-vocabulary words, artificially lowering the perplexity of closed-vocabulary LMs, particularly on data with a large number of OOVs.
Removing those same words in evaluating a sub-word language model, would disadvantage it.

A better alternative, which this paper will use, is to compare model
performance the word-level accuracy.
The most accurate way would be to find the word with the highest probability by summing over sequences of tokens.
However, we choose a simpler,
though less accurate method \citep[similar to][]{likhomanenko2019who}:
repeatedly generate the best tokens within each word's bounds and only
accept the word as accurate if all generated tokens were correct.

\subsubsection{Comparing sub-word with sub-word}

It is possible to meaningfully compare perplexities of two language
models with different sub-word tokenizations~\cite{Mie2016Can}.
Though the language model assigns probability mass to all token
sequences, a single sentence can have multiple corresponding
token sequences, only one of which will be chosen by the tokenizer.  Some of
the probability mass will therefore be lost to never-occurring token
sequences.  However, it is unfeasible to sum over all token sequences
\citep{likhomanenko2019who}.

The danger with comparing perplexities directly is
that since models with different tokenizers operate on different sets of
tokens the number of tokens needed to encode each sentence is different
in general \cite{Mie2016Can}.  Nevertheless, note that all models assign a
probability to a sentence (with the approximation above).
To compute the perplexity in such a way that it can be compared across tokenizers, use the same denominator in computing the
perplexity: the number of words in the sentence instead of number of
tokens, which depends on the tokenizer.  Therefore we define the
perplexity as:
\begin{equation}
  ppl_{\theta, \tau}(s) = \exp \left(\frac{-\log(P_{\theta, \tau}(s))}{\lVert s \rVert_w} \right)
  \label{eq:perplexity}
\end{equation}
where $\lVert s \rVert_w$ counts the number of words in the sentence
$s$.
To generalize from a single sentence to a dataset, replace $s$ with the concatenation of all sentences in the dataset.

\begin{figure*}[!t]
    \centering
    \includegraphics[width=1.0\linewidth]{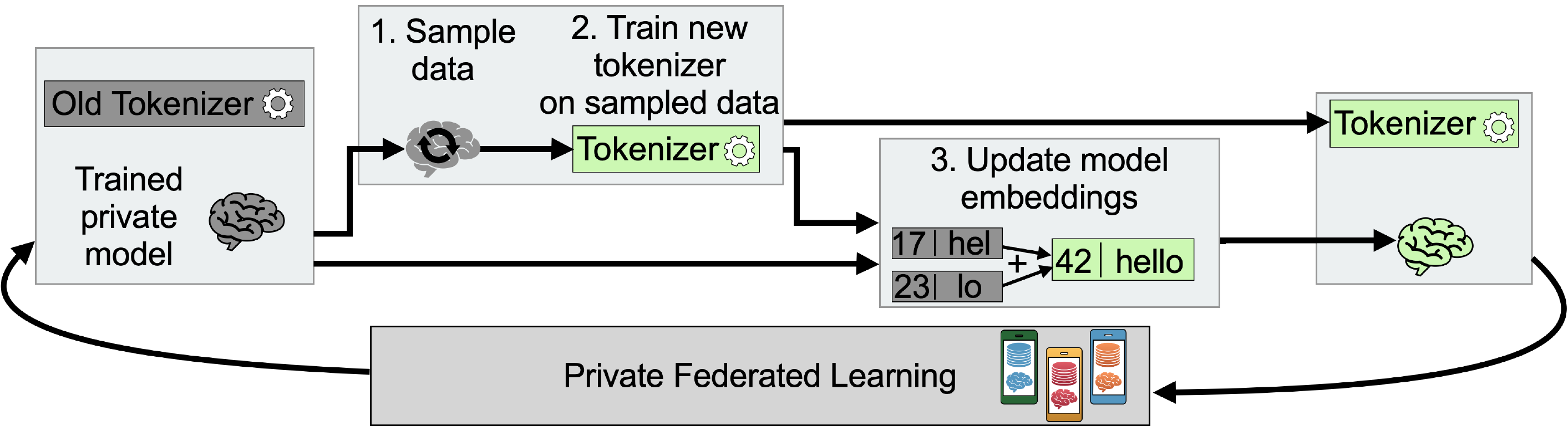}
    \caption{New pipeline for updating the tokenizer through model sampling.}
    \label{fig:pipeline}
\end{figure*}

\section{Learning a Tokenizer with Private Federated Learning}

\paragraphbe{Problem definition.} We aim to obtain a tokenizer that
works well on users' federated data without compromising user
privacy. First, we aim to find the appropriate tokenization scheme, and
second, given the tokenization scheme obtain the right approximation of
user data to train the tokenizer.

\paragraphbe{Setting} We focus on a common application of federated
learning: training a language model, parameterized by $\theta$, using
federated learning with differential privacy. In our setting each user
$u_i$ has a dataset $d_i$ of private texts from a private distribution
of user data $\mathcal{D}$.  The trained model will be evaluated against
a held-out dataset $\mathcal{D}_{test}$, e.g.\ a mix of all user data,
which in practice must be replaced by federated evaluation.

We assume that the central server does not have access to the user data
distribution $\mathcal{D}$ and can only approximate it with the publicly
available dataset $\mathcal{D}_{pub}$.  We assume the public data is
some commonly available dataset, such as Wikipedia
\cite{merity2016pointer}.  The tokenizer trained on this public data
will be $\tau_{pub}$.  For comparison we assume the existence of an
\emph{oracle} tokenizer $\tau_{o}$ initialized on users' training data
$\mathcal{D}$.

Papers that study language models in federated learning commonly use
word-level tokenization.  While some papers \citep[e.g.][]{fedlearn_dp},
build the vocabulary using publicly available dataset, others
\citep[e.g.][]{yu2020salvaging, caldas2018leaf} explicitly use the
federated training data, even though in real-world scenarios the
analogous data would be unavailable and it violates privacy guarantees
when used in PFL \cite{li2021ditto}.

\subsection{Sampling from a PFL-trained language model}

To address the problem of learning a good tokenizer we first propose to
use a sub-word tokenizer with an open vocabulary. This allows the
language model trained with such a tokenizer to represent any word, if
inefficiently. It is then possible to query the language model to find
new words as the model can utilize this open vocabulary.  This is the
core of the Algorithm~\ref{alg:sampling} that this paper introduces.

Figure \ref{fig:pipeline} shows the proposed pipeline.  A language model
is trained with private federated learning.  This results (on the left)
in a model matched with an old, stale tokenizer.  The next block queries the
language model to produce a better tokenizer, with a method that section
\ref{sec:sampling} will detail.  The block after that updates the
language model for the new tokenizer, using reasonable guesses for the
new parameters.  This results in a new LM-tokenizer combination that can
be trained further with PFL.

We assume that the language model obtained with the stale tokenizer is
trained with a certain privacy budget.  The postprocessing guarantee of
differential privacy~\cite{dwork2011differential} means that the steps
other than private federated learning do not consume any further budget.
The function \textsc{Update} in Algorithm~\ref{alg:sampling} performs
the on-server steps.  The following sections will give more detail.

\subsection{New tokenizer from a trained LM}
\label{sec:sampling}

Training a tokenizer requires text data.  Since the raw data is not
available, we propose to instead sample from the LM matched with the
stale tokenizer, as detailed in Algorithm~\ref{alg:sampling}.  The
\textsc{SampleTokens} function samples from the language model, drawing
sequences of tokens according to the probabilities that the model
assigns to them.  The \textsc{Sample} function then converts these
sequences in the old tokenization into word sequences, by decoding with
$\tau_{pub}$.  Once a large enough corpus of word-level sentences has
been produced, training a tokenizer proceeds as normally (the
\textsc{TrainTokenizer} function is not specified).

\newcommand{\doubleplus}{+\!\!\!+\,}

\subsection{Adapting the language model to the new tokenizer}
\label{sec:change_tokenizer}

After a new tokenizer $\tau$ has been trained, the language model,
trained with $\tau_{pub}$, must be updated to work with the new
tokenizer.  Neural-net language models use an embedding layer to convert
the provided tokens into multi-dimensional vectors.  It is the embedding
vectors that are most important to modify when changing the
tokenization.  The rest of the model only consumes the embedding vector.
It is not possible to find the optimal parameters without further
training of both embeddings and other layers, but we propose an
algorithm to find a reasonable starting point, in the function
$\text{\textsc{Remap}}(\tau, \tau_{pub})$ in
Algorithm~\ref{alg:sampling}.

\textsc{Remap} iterates over the tokens from the new tokenizer $\tau$
and creates the mapping from the tokens' embedding in the public
tokenizer $\tau_{pub}$ to the new token's embedding. In some cases it is a one-to-one mapping, but
when the new token accumulates multiple tokens in $\tau_{pub}$ we split
the weight equally between each token.

Once we have the mapping $map$ we modify the embedding layer of the
model by performing matrix multiplication, i.e.\ $\theta.\mathrm{embedding} = map
\cdot \theta.\mathrm{embedding}$.  The resulting model can accept the tokens from
the new tokenizer $\tau$, and can participate in future training in
federated learning.

\begin{algorithm}[t]
    \caption{Model sampling algorithm}
    \label{alg:sampling}
    \begin{algorithmic}
    \State \textbf{\textit{Inputs:}} model $\theta$, current sentence $s$, new
    tokenizer $\tau$, public tokenizer $\tau_{pub}$, size of the sampled
    dataset $\mathrm{corpus\_size}$.
    \vspace{0.1cm}
    \Function{SampleTokens}{$\theta, s$}
        \State $t_{next} \sim_\theta t_k | s$
        \If {$t_{next} = \EOS$}
            \State \textbf{return} $s \doubleplus t_{next}$
        \Else
            \State \textbf{return} \textsc{SampleTokens}($\theta, s \doubleplus t_{next}$)
        \EndIf
    \EndFunction
    \vspace{0.1cm}

    \Function{Sample}{$\theta, \tau$}
        \State \textbf{return} $\tau.\mathrm{decode}($
        \State $\qquad \text{\textsc{SampleTokens}}(\theta, [\BOS]))$
    \EndFunction
    \vspace{0.1cm}

    \Function{Remap}{$\tau_{pub}, \tau$}
    \State $\mathrm{map} = \mathrm{zeros}(\tau.\mathrm{size}, \tau_{pub}.\mathrm{size})$
    \For{$\mathrm{token}, \mathrm{tid} \gets \tau.\mathrm{vocab}$}
        \State $\mathrm{tokens} = \tau_{pub}.\mathrm{decode}(\mathrm{token})$
        \For{$\mathrm{token} \gets \mathrm{tokens}$}
            \State $\mathrm{tid}_{pub} = \tau_{pub}.\mathrm{vocab}[\mathrm{token}]$
            \State $\mathrm{map}[\mathrm{tid}_{pub}, \mathrm{tid}] = 1/\mathrm{len}(\mathrm{tokens})$
        \EndFor
    \EndFor
    \State \textbf{return} $\mathrm{map}$
    \EndFunction

    \Function{Update}{$\theta, \tau_{pub}$}
    \While{$\mathrm{len}(\mathrm{corpus}) < \mathrm{corpus\_size}$}
        \State $\mathrm{corpus} \leftarrow \textsc{Sample}(\theta, \emptyset, l_{max})$
    \EndWhile
    \vspace{0.1cm}

    \State $\tau = \textsc{TrainTokenizer}(\mathrm{corpus})$
    \State $\mathrm{map} = \textsc{Remap}(\tau_{pub}, \tau)$
    \State $\theta.\mathrm{embedding} = \mathrm{map} \cdot \theta.\mathrm{embedding}$
    \State \textbf{return} $\theta, \tau$
    \EndFunction

    \end{algorithmic}
\end{algorithm}
\section{Experiments}
\label{sec:exps}

We evaluate our approach by first looking at performance of tokenizers
trained on the distributions matched and mismatched to real data, we
then test the proposed federated sampling  on different datasets for
federated learning.

\subsection{Experimental setup.}

We use two datasets common in the federated learning literature
\cite{kairouz2019advances}.  While both use English, there is nothing
about our experiments that is specific to this language, and multilingual datasets can further benefit from using SentencePiece tokenization~\cite{kudo2018sentencepiece},.  

\begin{itemize}
    \item Reddit data -- this dataset is taken from the LEAF benchmark
    \cite{caldas2018leaf} and contains over a million users that have
    multiple posts on the Reddit platform. As proposed by LEAF, we limit
    each user to contain at most 1600 tokens and use 10\,\% of users for
    faster training.
    \item StackOverflow data -- this data is taken from Kaggle
    \cite{stackoverflow} and processed with the TensorFlow Federated
    framework. The train split of the dataset contains 342k users and we
    select at most 1600 tokens per user.
\end{itemize}

\paragraphbe{Model parameters.} We use an LSTM model with 3 layers, and
total parameters of 14M.  We also use a Transformer language model~\cite{vaswani2017attention} with
6 layers and the same total number of parameters as the LSTM (see Appendix~\ref{sec:ablation}).  Each
model is trained from scratch.

\paragraphbe{Hyper-parameters.} 
We set the privacy budget to $\epsilon=2$ and $\delta=10^{-6}$ -- a common privacy regime~\cite{kairouz2019advances}.
For the ``heavy hitters'' baseline we use local DP with an additional privacy budget of $\epsilon=8$.%
\footnote{Budgets for local and central privacy are not immediately comparable, but see \citet{feldman2021hiding}.}
The overall population for the
moments accountant is assumed to be 10m.  We use a cohort size of
$20,000$ for each round and train all models for $5,000$ iterations.  We use
Adam~\cite{kingma2014adam} for central optimization with learning rate
set to 0.5.  For the clients we use SGD and train for $1$ local epoch with
batch size set to 16 and local learning rate set to 0.1, and an $L_2$ clipping bound for DP of $0.5$.

\paragraphbe{Vocabulary size.} We assume that the tokenizer has a
moderate vocabulary size such as 10,000 tokens (we experiment with
larger vocabularies in Appendix~\ref{sec:ablation}). Smaller
vocabularies reduce model size and, therefore, might be better for
deployment on devices and communication with the global server.

\paragraphbe{Tokenizer details.} To train an initial tokenizer we use a
popular and public Wikipedia dataset \cite{merity2016pointer}.  It may
seem like the distribution of Wikipedia data is artificially far from
the distributions of Reddit and StackOverflow data. However, the server
might not have the right prior possibly due to a natural
\emph{distribution shift}~\cite{miller2020effect} of typed texts (such
as an emerging topic of which there were plenty recently).

We use BPE and WordLevel tokenization algorithms from the HuggingFace
Tokenizer library \cite{huggingfacetok}.  Each user post is surrounded
by special tokens {\BOS} and {\EOS}.  We also tried WordPieces
tokenization which has slightly better performance than BPE but cannot
encode all words and is therefore less applicable in FL.

\paragraphbe{Note on
splitting data.}   Whereas the original LEAF dataset for Reddit proposes
to split each user's data we argue that in real life not every user might
have a chance to participate in the training.  Therefore, we split users
into two distinct training and test sets and evaluate the model on data
from the users who have never participated in the training. This results
in notably increased test perplexity but provides a clear separation
between training and inference modes. 

\begin{table}[t!]
	\centering
	\footnotesize
	\caption{Word accuracy suffers for word-level tokenization that uses mismatched data.}
	\label{tab:word_level}
	\begin{tabular}{ll|r@{~~}@{~}r@{~~~~}r@{~}}
		& & \multicolumn{2}{c}{$\tau$ statistics} & Word \\
		Type & Data  & \OOV & Tokens & Accuracy \\
		 & to train $\tau$  & (\%) & per word & (\%)  \\
		\midrule
		\multicolumn{5}{c}{\vspace{0.2cm}\textit{Reddit}} \\
		Word-Level & Wiki
             & 13.0 & 1.00     & 17.7  \\
             \vspace{0.2cm}Word-Level & Oracle
             &  5.5 & 1.00     & 24.1 \\
		BPE & Wiki
             & 0.0 & 1.32 & 22.2  \\
		BPE & Oracle
             & 0.0 & 1.22 & 22.5 \\
        \midrule
		\multicolumn{5}{c}{\textit{StackOverflow}} \vspace{0.2cm}\\
		Word-Level & Wiki
            &  9.8 & 1.00     & 30.0  \\
		\vspace{0.2cm}Word-Level & Oracle
            &  2.0 & 1.00     & 33.0\\

            BPE & Wiki
             & 0.0 & 1.41 & 31.8   \\
		BPE & Oracle
             & 0.0 & 1.24 & 32.4 \\
		\bottomrule
	\end{tabular}
\end{table}

\subsection{Comparing tokenization schemes}
\label{sec:comparetok}

Table~\ref{tab:word_level} summarizes experiments that use different
tokenization schemes.
We compute statistics on tokenizers: the average share of \OOV tokens for the
word-level scheme and the average number of tokens required to encode one
word for the sub-word scheme.
To compare the effect of each tokenizer on the PFL-trained model, we report word-level accuracy, for the reasons described in Section~\ref{sec:compare_tokenizations}.
The ``wiki'' tokenizers are trained on the
Wikipedia data, and the ``oracle'' tokenizers directly on the training
data.

Word-level tokenization provides high word accuracy when it
is trained using ``oracle'' user training data. However, when the
word-level has access to only public ``wiki'' dataset that mismatches
user distribution the performance
significantly drops: by 26\,\% for Reddit and 10\,\% for StackOverflow
with a significant increase in out-of-vocabulary share. However, BPE
tokenizers that use public data perform more consistently and outperform the
word-level models trained on public data, but still require a large number
of tokens per each word.

\subsection{Learning a tokenizer with sampling}
\label{sec:expsampling}

A key part of the proposed algorithm is the sampling from a model that
uses a public tokenizer $\tau_{pub}$, but is trained with private
federated learning and should represent the words in the actual
data.  The sampling is implemented as in Algorithm \ref{alg:sampling}.

\begin{figure}[b!]
      \centering
      \begin{minipage}{0.85\linewidth}
          \raggedright
          {\small \emph{Reddit}}

          {\footnotesize i would love to know why we may already live in a consolation subreddit and the aforementioned it will almost always be done on the warrior sheet shows from the west . i}

          ~

          {\small \emph{StackOverflow}}

          {\footnotesize json results are : can anyone provide a complete sample response ( lists of descendants list ) to my page depending on future python functions . in web apps that require patient for many}

      \end{minipage}
      \caption{Example of sampling data from the model.}
      \label{fig:sampling_example}
  \end{figure}

First, Figure \ref{fig:sampling_example} shows samples from the language
models on the two data sets.  Although clearly the samples are less
coherent than the underlying data, it seems plausible that the word
occurrences match that data.

\begin{table}[t!]
	{\centering
	\footnotesize
	\caption{Tokenizers initialized on sampled data perform very close to using ``oracle'' data.}
	\label{tab:main}
	\begin{tabular}{l@{~~~}l@{~}|r|r|r@{~~~~~}r}
		& &  &  & \multicolumn{2}{c}{LM} \\
		Type &  Data & Data & Tokens & Acc. & Perp. \\
		       &  to train $\tau$ & KLD  & p/word & (\%) & \\ 
		\midrule
		\multicolumn{5}{c}{\textit{Reddit}} \\[0.2cm]
		BPE & Wiki
            & 0.78 &  1.32 & 22.2 & 276.5  \\
		BPE & Oracle
            & 0 &  1.22 & 22.5 & 256.9 \\[0.2cm]
		BPE & Heavy hitters$^*$
            & 0.09 &  1.30& 22.1& 274.2  \\
            BPE & \textbf{Sampled}
            & 0.02 &  1.22 & 22.5 & 257.7  \\
		\midrule
		\multicolumn{5}{c}{\textit{StackOverflow}} \\[0.2cm]
		BPE & Wiki
            & 1.06 &1.41 & 31.8 & 124.6  \\
		BPE & Oracle
            & 0 &  1.24 & 32.4 & 108.2  \\[0.2cm]
            BPE & Heavy hitters$^*$
            & 0.10 &  1.29 & 32.1 & 115.9 \\
		BPE & \textbf{Sampled}
            & 0.01 & 1.23 & 32.4 & 108.7  \\
		\bottomrule
	\end{tabular}
      }
      {\small
      $^*$The ``heavy hitters'' algorithm requires additional privacy budget.}
\end{table}

\begin{figure*}[t!]
      \subfigure[{Reddit dataset}]{
            \includegraphics{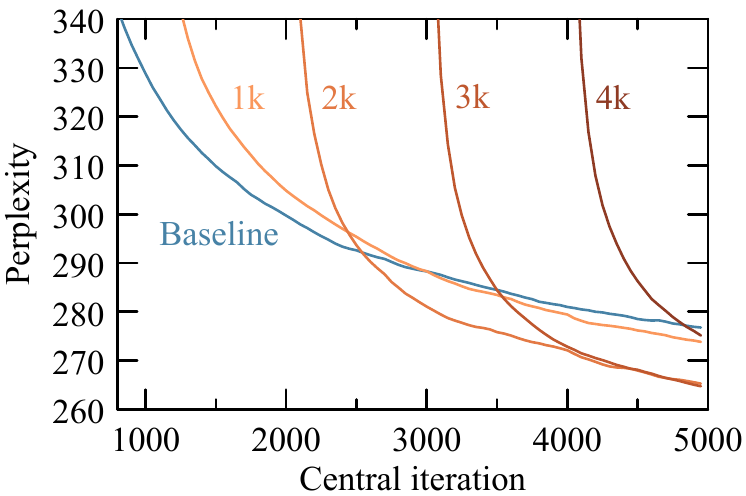}}
      \hspace{\stretch{1}}
      \subfigure[{StackOverflow dataset}]{
            \includegraphics{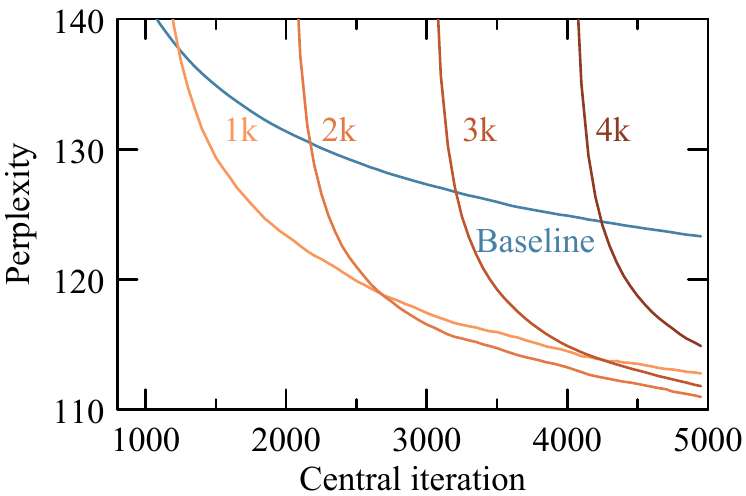}}
      \caption{Perplexity for switching the tokenizer at different rounds of federated learning.}
	\label{fig:iterations}
\end{figure*}

Second, Table~\ref{tab:main} further investigates the properties of the
sampled text.  The ``BPE sample'' rows refer to the method proposed in
this paper.  A language model with the ``wiki'' tokenizer is trained
with PFL on the first half of the training data.  Then samples are drawn
from this language model.  Then, the language model is trained from
scratch on the second half of the training data.

The ``BPE Heavy hitters'' rows refer to training with a differentially private
``heavy hitters'' algorithm \cite{apple2017learning}.  Each of the
population of the users from the first half of the
training set contributes three words from the from
the Wikipedia dataset, with a local privacy budget of $\epsilon=8$.
Just like for the sampling approach, the language model is then trained
from scratch on the second half of the training data.

First, we examine the difference between the real training data and the
data used to train the tokenizers.  The column ``Data KLD'' shows the KL
divergence from the user ``oracle'' training data to the sampled data.  The KL
divergence is computed from the unigram counts, which are relevant for
training a tokenizer, over the top
10,000 words from the training data and with add-1 smoothing.  The KL divergence to the training
data itself, which the oracle tokenizer is trained on, is 0 by
definition.  The KL divergence between the actual data and the Wikipedia
data, on the other hand, is around 1, for both datasets.  Both the heavy
hitters algorithm and the algorithm we propose in this paper find a
distribution close to the real distribution.

For sub-word tokenizers, the number of tokens per word is relevant.
Even though they can represent unseen words by multiple tokens, a
language model trained on top of that has a harder task given the
longer context on average.  The oracle tokenizer has the lowest number
of tokens per words and the ``wiki'' tokenizer the highest.  The ``BPE
sample'' tokenizer comes very close to the oracle tokenizer. 

However, the heavy hitters experiment shows much smaller gain in
performance, i.e. better than ``wiki'' tokenizer but still worse than
our proposed sampling method. Furthermore, it requires a separate
privacy budget allocated for the run, while sampling can operate on
existing prior model.

\subsection{Iterative updates}

This part implements Algorithm \ref{alg:sampling} completely.  We
again initialize the tokenizer on publicly available data.  We then
train the language model with PFL.  At a point during training, we
retrain the tokenizer by sampling.
Unlike in the previous section, we
update the language model by remapping its embedding layer, and continue
training.  We sample the same data before and after changing the
tokenizer.

Figure~\ref{fig:iterations} shows the results for changing tokenizers at
different times.
The ``Baseline'' curve represents the model trained using public
tokenizer $\tau_{pub}$ from Wikipedia data.
Each of the other curves takes the system from the ``Baseline'' curve at a
different iteration.  As expected, the initial remapping of the
embedding layer is not perfect and needs finetuning.  The graph also
shows the tradeoff in when to change tokenizers: too early, e.g.\ after
only 1000 iterations, and the tokenizer is not representative enough
yet; too late, e.g.\ after 4000 iterations, and there is not enough time
to converge again.

\section{Conclusion}

This paper has proposed a method that allows a tokenizer to be found together with a language model using private federated learning.
First, it has shown that a mismatched tokenizer can cause a significant performance degradation.
The key to improving this is to use a sub-word tokenizer which allows new words to be represented as a sequence of tokens.
Then, a language model trained with PFL can represent the private data.
This paper has presented a method to produce a new tokenizer from that model, and to convert the model to work with the new tokenizer.
When this is trained further with private federated learning, it outperforms the language model with the mismatched tokenizer, and gets close to one with the oracle tokenizer.

\paragraphbe{Personalization and Fairness.}
The problem of out-of-vocabulary words might be more acute for some users that use unique vocabulary, such as dialect, and impact individual performance.
Therefore good tokenizers can benefit personalization in federated models \cite{li2021ditto,yu2020salvaging}.



\bibliography{anthology,main}
\bibliographystyle{acl_natbib}

\clearpage
\appendix
\section{Impact of hyperparameters}
\label{sec:ablation}

\begin{figure}
	\centering
	\includegraphics{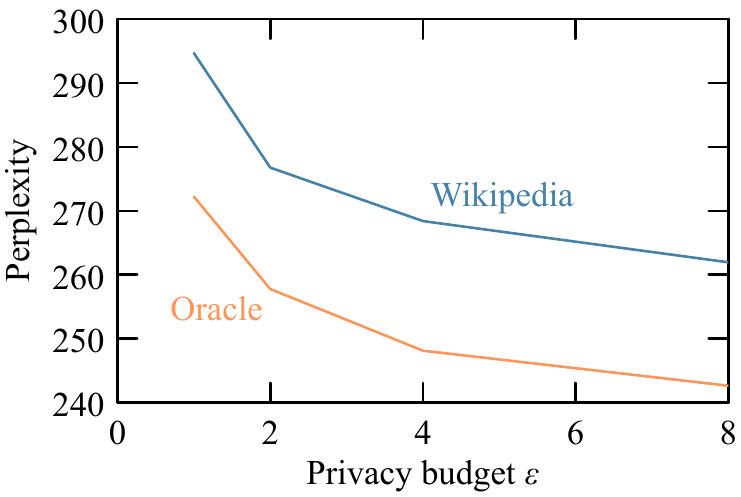}
	\caption{Perplexity trained with different privacy parameter $\epsilon$.}
	\label{fig:privacy_params}
\end{figure}

\begin{figure}[t]
	\centering
	\includegraphics{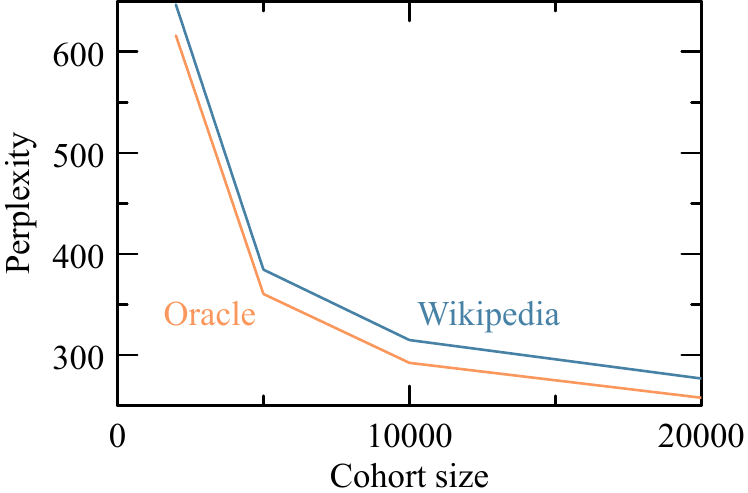}
	\caption{Perplexity trained with different cohort sizes.}
	\label{fig:cohort_size}
\end{figure}

This section examines different hyperparameters.

\subsection{Experimental design}

First, consider the choice to train the public tokenizer on Wikipedia data.
To examine the effect of using a more conversational style corpus.
To do this, Table \ref{tab:wikipedia} takes a subset of the numbers from Table \ref{tab:main} and adds a scenario where a tokenizer on StackOverflow data is used with Reddit data and vice versa.
The cross-dataset numbers are highlighted bold in the table.

First, in terms of the KL divergence the StackOverflow data seems a slightly better model for the Reddit distribution than the Wikipedia data is.
However, when using PFL to train on Reddit data, but with a StackOverflow-trained tokenizer, the perplexity deteriorates compared to the Wikipedia-trained tokenizer.
Second, the reverse experiment looks a bit better but not hugely better.
Though the KL divergence from the StackOverflow data to the Reddit data is significantly better than the KL divergence to the Wikipedia data, some of that advantage disappears in the final trained model.

\begin{table}
	\centering
	\caption{The effect of using the Wikipedia corpus against the results in Table~\ref{tab:main}.}
	\label{tab:wikipedia}
	\begin{tabular}{ll|@{~~}l@{~~}|@{~~~}c}
		\toprule
		$\tau$ &  Data & Data & LM \\
		       &       & KLD  & perp.\\
		\midrule
		\multicolumn{4}{l}{\textit{Reddit}} \\
		BPE & Wikipedia
            & 0.7826 & 276.5  \\
		BPE & \textbf{StackOverflow}
            & 0.6046 & 283.6  \\
		BPE & Reddit
            & 0      & 256.9  \\
			\midrule
			BPE & sample
            & 0.0212 & 257.7  \\
		\midrule
		\multicolumn{4}{l}{\textit{StackOverflow}} \\
		BPE & Wikipedia
            & 1.0629 & 124.6  \\
		BPE & \textbf{Reddit}
            & 0.5315 & 118.8  \\
		BPE & StackOverflow
            & 0      & 108.2  \\
		\midrule
		BPE & sample
            & 0.0089 & 108.7  \\
		\bottomrule
	\end{tabular}
\end{table}

Then, consider the choice of vocabulary size, here the number of distinct tokens.
Table \ref{tab:vocabsize} shows the perplexities for the baseline (``Wiki'') and ceiling (``oracle'') experiments.
Though the absolute numbers change, the trends do not change.

\begin{table}
	\centering
	\caption{The effect of varying the vocabulary size.}
	\label{tab:vocabsize}
	\begin{tabular}{l|rr|rr}
		\toprule
		Vocab  size &\multicolumn{2}{c|}{Reddit} & \multicolumn{2}{c}{StackOverflow} \\
		&Wiki &  Oracle  &Wiki &  Oracle \\
		\midrule
		5,000 & 304.3 & 282.2 & 136.3 &  116.8 \\
		10,000 & 276.5 & 256.9  & 124.6 &  108.2 \\
		50,000 & 243.9 & 225.4    & 111.5 & 101.5 \\
		100,000 & 231.2 & 217.9   & 108.9 &  100.5 \\
		\bottomrule
	\end{tabular}
\end{table}

Similarly for changing model architectures.
This paper has presented results on an LSTM model.
Table \ref{tab:modelarch} shows results on a Transformer model.
Again, though the absolute numbers change, the trends do not change.

\begin{table}
    \centering
    \caption{The effect of changing model architectures.}
    \label{tab:modelarch}
    \begin{tabular}{l|rr|rr}
        \toprule
        Model &\multicolumn{2}{c|}{Reddit}& 
        \multicolumn{2}{c}{StackOverflow}\\
     architecture &Wiki &  Oracle  &Wiki &  Oracle \\
     \midrule
     Transformer  & 261.9 & 244.8 & 117.4  & 107.0 \\
     LSTM  & 276.5 & 256.9 & 124.6  & 108.2  \\
     \bottomrule
    \end{tabular}
\end{table}

\subsection{Other hyperparameters}

We consider two hyperparameter choices for experiments: first, the privacy budget, and secondly, the cohort size.

Figure \ref{fig:privacy_params} shows the effect of different privacy parameters.
The effects are not huge, but clearly differential privacy does impede learning somewhat.

Figure \ref{fig:cohort_size} shows the effect of differing cohort sizes.
A larger cohort size implies a better signal-to-noise ratio when training with differential privacy.
However, for practical reasons it is preferable for cohorts to be smaller.
10,000 is a happy medium between good performance and practicality.
Also, again, though the absolute numbers change, the trends do not change.

\end{document}